\documentclass[%
reprint,
superscriptaddress,
amsmath,amssymb,
aps,
prl,
]{revtex4-2}

\usepackage{graphicx}
\usepackage{dcolumn}
\usepackage{xcolor}
\usepackage{ulem} 
\usepackage{bm}
\usepackage{multirow}
\usepackage{array}
\usepackage{hyperref}
\usepackage{float}
\newcommand{\iitj}{\affiliation{Polymer Electrolytes and Materials Group (PEMG), Department of Physics, Indian Institute of Technology Jodhpur\\
		N.H. 62, Nagaur Road, Karwar, Jodhpur, Rajasthan, India - 342030. }}
\newcommand{\squ}{\affiliation{Currently on a sabbatical at the Department of Physics, Sultan Qaboos University, PO Box 36, Al-Khoud 123, Muscat 123, Oman}}
\makeatletter
\renewcommand{\fnum@figure}{Figure~\thefigure}
\makeatother


\begin{document}
	
	\title{A hybrid Green-Kubo (hGK) framework for calculating viscosity from short MD simulations}
	\preprint{APS/123-QED}
	
	\author{Akash K. Meel$^\dagger$}
	\iitj

	\author{Santosh Mogurampelly$^*$}
	\iitj
    \squ
	
	
	
	
	\date{\today}
	
	\begin{abstract}
        \begin{center}
        \textbf{Abstract}
        \end{center}
		Viscosity calculation from equilibrium molecular dynamics (MD) simulations relies on the traditional Green-Kubo (GK) framework, which integrates the stress autocorrelation function (SACF) over time. While the formalism is exact in the linear response regime, the traditional approach often suffers from poor convergence and requires extensive phase space sampling, which is computationally demanding for soft matter and polymer systems. In this Letter, we introduce a hybrid Green-Kubo (hGK) framework that alleviates these limitations by partitioning the SACF into two physically meaningful regimes: (i) a short time ballistic component extracted directly from short MD simulations, and (ii) a long time relaxation tail represented using analytically motivated functions, $\phi(\tau)$, fitted only to short trajectories such that $\eta = \frac{V}{k_B T} \left[\frac{1}{6} \sum_{\alpha\beta}\int_{0}^{\tau_l} \langle P_{\alpha\beta}(t) P_{\alpha\beta}(t+\tau) \rangle \, d\tau\;+\; \int_{\tau_l}^{\infty} \phi(\tau)\, d\tau \right]$. This strategy bypasses the need for extensive sampling while preserving the exact GK framework at short times. Benchmarking against SPC/E water confirms excellent agreement with established results, and we further demonstrate the efficacy of the method for challenging electrolyte systems (EC-LiTFSI and PEO-LiTFSI), for which the GK framework fails to converge. The computational savings are substantial, with reductions of several orders of magnitude in required sampling, achieved without compromising predictive accuracy. We also discuss the limitations of the hGK framework and outline clear avenues for refinement, including optimal tail selection and robust identification of relaxation regimes in noisy stress data. The hGK framework presented in this Letter provides a conceptually simple, broadly applicable, and computationally efficient route for viscosity prediction in molecular liquids, polymer melts, and ionically conducting soft materials.
		\begin{description}
		\item[Authors to whom correspondence should be addressed]$^\dagger$\hyperlink{p24ph0002@iit.ac.in}{p24ph0002@iit.ac.in}, and $^*$\hyperlink{santosh@iitj.ac.in}{santosh@iitj.ac.in}
		\end{description}
	\end{abstract}

\maketitle



Viscosity is one of the key transport coefficients of fluids and soft matter, characterizing the resistance to shear deformation\cite{bird2006,hansen2013}. This macroscopic property originates from molecular processes that include collisional transfer, fluctuations of the momentum flux, and correlated many body rearrangements\cite{pathria2021,kirkwood1946,mori1958}. Apart from the fundamental scientific point of view, understanding viscosity is essential for designing paints, lubricants, food products, polymer melts, and electrolyte solutions\cite{keane2025,zhao2016,sharma2002,asbeck1954}. Experimental measurements using capillary viscometers, rotational rheometers, and oscillatory devices provide robust benchmarks\cite{suman2025,watson1902,gaeta1973,sikdar1979}, but molecular dynamics (MD) simulations offer a complementary route that links macroscopic viscosity to microscopic structure and relaxation pathways\cite{frenkel1996,allen2017}. MD enables prediction before synthesis and provides deeper insights into the underlying mechanisms that govern momentum transport.

Viscosity in MD is commonly obtained through nonequilibrium approaches or through equilibrium simulations. Nonequilibrium molecular dynamics (NEMD) induces shear flow and extracts viscosity from the resulting stress response\cite{ciccotti1979,todd2017}. Classic implementations of NEMD often use the SLLOD equations of motion with Lees-Edwards boundary conditions\cite{evans1984,lees1972}, and a careful treatment of the long range electrostatics is essential in polar liquids\cite{wheeler1997}. Among other methods, the reverse NEMD employs momentum exchange to generate a steady velocity gradient\cite{muller1999}. These methods are powerful but require careful control of shear rates and system size effects.

\begin{figure*}[t]
    \centering
    \includegraphics[width=6.75 in, keepaspectratio]{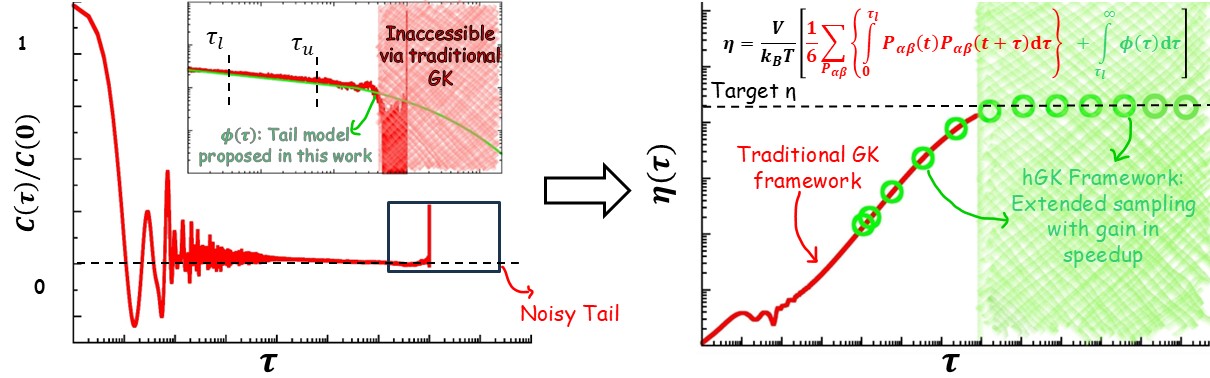}
    \caption{Schematic illustration of the hGK framework. Left: A representative SACF showing an oscillatory short time region followed by a noisy long time tail. Only data up to $\tau_u$ are reliably sampled; beyond this window, the SACF is replaced by a fitted tail function $\phi(\tau)$ that captures the slow relaxation inaccessible to the traditional GK framework. Right: Running integral of viscosity $\eta(\tau)$ from the traditional GK framework, which fails to plateau, compared to the smooth convergence obtained using hGK. Tail reconstruction recovers the missing long time contribution and enables reliable viscosity calculation from short trajectories.}
    \label{fig:figure1}
\end{figure*}

Equilibrium molecular dynamics (EMD) provides an alternative based on the traditional Green-Kubo (GK) framework\cite{green1954,kubo1957,kubo1966}, which connects relevant macroscopic transport coefficients to underlying microscopic fluctuations\cite{onsager1931,onsager1936}. In this formulation, viscosity is obtained from the time integral of the stress autocorrelation function (SACF)\cite{zwanzig1965,hess2002}. The SACF captures the relaxation of collective stress fluctuations: its short time decay reflects rapid molecular motions, whereas its long time behavior is governed by slower structural relaxations or local molecular rearrangements\cite{callen1951,evans2008}. Short MD trajectories are sufficient to resolve the rapid decay of the SACF, but many complex liquids and soft materials exhibit slow relaxation that persists well beyond accessible simulation times. Accurate evaluation of the GK integral then requires a large number of uncorrelated and long trajectories for extensive ensemble averaging, which becomes increasingly impractical for complex fluids and most soft matter systems. As a result, the SACF tail is dominated by statistical noise at large lag times, and the running integral fails to reach a stable plateau. This slow and noisy decay of the SACF is the primary limitation of the GK framework.

To address this challenge, we propose a hybrid Green-Kubo framework (hGK) that extracts the short time SACF from MD and replaces its long time tail with an analytic continuation as presented in Figure \ref{fig:figure1}. The key idea is to capture the physically meaningful relaxation regimes from short trajectories that adequately represent the true phase space. In the hGK framework, the SACF is written as a sum of two contributions: a short time component taken directly from simulation and a long time component modeled with an analytical function, $\phi(\tau)$, fitted only to short segments of the SACF. The running viscosity is then written as
\begin{align}
    \eta = \frac{V}{k_B T} \Bigg[\frac{1}{6} \sum_{\alpha\beta}\int_{0}^{\tau_l}\langle P_{\alpha\beta}(t) P_{\alpha\beta}(t+\tau) \rangle\,d\tau\notag\\
    +\int_{\tau_l}^{\infty} \phi(\tau)\, d\tau\Bigg],
\end{align}
where $P_{\alpha\beta}$ for $\alpha$ and $\beta=x,y,z$ are components of the stress tensor, $V$ is the volume, $T$ the absolute temperature, and $k_B$ is the Boltzmann constant. In this context, the SACF is defined as $C(\tau) = \langle P_{\alpha\beta}(t)P_{\alpha\beta}(t+\tau)\rangle$, with ensemble averaging taken over off diagonal elements of the stress tensor, independent EMD trajectories and all time origins. Increasing the number of independent trajectories and time origins reduces statistical noise in the SACF, particularly at long lag times where fewer time origins contribute. As a result, the uncertainty in the viscosity estimate is controlled by the statistical quality of the SACF in the short time regime where direct sampling is reliable. The hGK framework exploits this property by restricting direct numerical integration to the statistically well sampled region and replacing the noise dominated long time tail with an analytical continuation. This formulation avoids the requirement of prohibitively long trajectories and extensive ensemble averaging at large lag times. It is worth noting that, although the long time tail extrapolation is model dependent, the short time correlations are still directly obtained from the EMD trajectories, consistent with the original GK framework. 

The concept of extracting reliable long time information from short simulations is reminiscent of trajectory extending kinetic Monte Carlo (TEKMC) strategies used in diffusion studies\cite{neyertz2010,hanson2011,mogurampelly2015,tien2018,dasgupta2023}. Neyertz and Brown demonstrated that short MD segments can be assembled into a Markovian model that reproduces long time dynamical behavior in glassy polymers\cite{neyertz2010}, providing access to timescales beyond the reach of direct EMD. Conceptually resembling TEKMC, the hGK adopts an analogous extrapolative approach to the SACF: the long time decay of the stress correlations is inferred by fitting analytical forms to well resolved short time data. Related ideas have appeared in the literature. Guo et al.\cite{guo2002} modeled the SACF of supercooled water using combined fast oscillatory and slow stretched exponential components, while Maginn et al.\cite{zhang2015} introduced a time decomposition procedure that reduces variance through block averaging. At a more fundamental theoretical level, Zaccone and co-workers\cite{zaccone2023,zaccone2025} have recently developed a nonaffine response formalism in which viscosity is computed directly from the vibrational spectrum of the liquid or melt, highlighting the challenge posed by long-time stress relaxation. These studies illustrate the value of extending equilibrium time correlations by combining short MD trajectories with physically motivated models, which aligns with the strategy used in the hGK framework. Along the lines of these ideas, the hGK framework integrates time decomposition and analytical continuation into a single operational procedure that avoids fitting noise-dominated correlation data.

The choice of the fitting window for accurately extracting $\phi(\tau)$ from short MD is central to the hGK framework. We identify a lower cutoff time $\tau_l$ as the point where the short time oscillations subside and the slow tail begins (see \textbf{Section S2} of \textbf{SI}). The present criterion used to identify $\tau_l$ is system specific and acts as a rule of thumb rather than a generalisable rule. Choosing $\tau_l$ too early forces the fit to reproduce rapid oscillations, while choosing it too late discards meaningful contributions to viscosity. Once $\tau_l$ is determined, we vary the upper limit $\tau_u$ and examine the predicted viscosity as a function of the window length $\tau_\Delta=\tau_u-\tau_l$. The predicted viscosity typically reaches a plateau as $\tau_u$ exceeds a threshold value, providing the shortest trajectory segment sufficient for a reliable calculation of viscosity. The plateau behavior with respect to  $\tau_\Delta$ is therefore used as an internal consistency check rather than an independent validation of the viscosity estimate. In addition, the structure of the stress autocorrelation function itself provides practical indicators of the applicability of the present formulation. Systems exhibiting secondary relaxation features, such as shoulders or multiple decay regimes in the SACF, may not be adequately described by a single stretched exponential continuation. Persistent systematic deviations between the analytical continuation and the SACF therefore indicate the presence of additional relaxation processes and may require extensions of the tail model beyond the present framework.

Another important aspect in optimizing the hGK framework is the selection of a suitable functional form for $\phi(\tau)$ when fitting the long tail region of the SACF. To this effect, we employ analytical functions that capture typical relaxation behaviors, such as power-law or simple exponential decays. For systems exhibiting multiple relaxation processes, it may be advantageous to include a series of decays in the tail model. In this work, however, we limit the analysis to cases where a single collective relaxation sufficiently describes the SACF, thereby avoiding the introduction of additional system specific parameters. In addition to power law and simple exponential forms, we employ a stretched exponential\cite{kohlrausch1854,williams1970},
\begin{align}
    \phi(\tau)=a_0\exp{\left[-\left(\frac{\tau}{\tau_*}\right)^\beta\right]},
\end{align}
a form well suited for liquids and polymer electrolytes where collective relaxation of SACF is broad and nonexponential\cite{lukichev2019,sturman2003}. The parameters $a_0$, $\tau_*$, and $\beta$ are obtained from nonlinear least squares fitting over the interval $[\tau_l,\tau_u]$. With this form, the viscosity can be written as the sum of the running integral of the short time contribution and an analytically integrated long time tail,
\begin{align}
    \eta(\tau') = \frac{V}{k_B T} \Bigg[\frac{1}{6} \sum_{\alpha\beta}\int_{0}^{\tau_l}\langle P_{\alpha\beta}(t)\,P_{\alpha\beta}(t+\tau)\rangle d\tau\notag\\
    +\int_{\tau_l}^{\tau'}a_0 \exp\left[-\left(\frac{\tau}{\tau^{*}}\right)^{\beta}\right] d\tau\Bigg].
\end{align}
This analytic continuation extends the viscosity integral smoothly to higher lag times and produces stable plateaus even when the SACF from traditional GK calculations fails to converge.

We evaluated the performance of the hGK framework for three representative systems spanning viscosities from $10^{-1}$ to $10^{3}$ mPa$\cdot$s: SPC/E water, liquid EC-LiTFSI, and polymeric PEO-LiTFSI. Water serves as a well characterized benchmark, while the two electrolytes represent technologically relevant liquid and polymer electrolyte systems with increasingly slow stress relaxations. Together, these systems provide a stringent test of the ability of hGK to extract long time behavior from short trajectories.

Water was modeled using the SPC/E three site rigid model\cite{berendsen1987}, and the EC-LiTFSI and PEO-LiTFSI systems were parameterized using the OPLS-AA force field\cite{jorgensen1996}. Initial configurations were generated with Packmol\cite{martnez2009}, followed by energy minimization and a two stage equilibration protocol. All systems were equilibrated in the NVT ensemble and then relaxed in the NPT ensemble at 1 bar to obtain the correct equilibrium density. Production simulations were carried out in the NVT ensemble at the target temperatures to accurately capture microscopic stress correlations. For each system, 50 independent trajectories of 10 ns were generated with a trajectory saving frequency of 1 fs, corresponding to a total sampling time of 500 ns per system.

\begin{figure}[t]
    \centering
    \includegraphics[width=3.375 in, keepaspectratio]{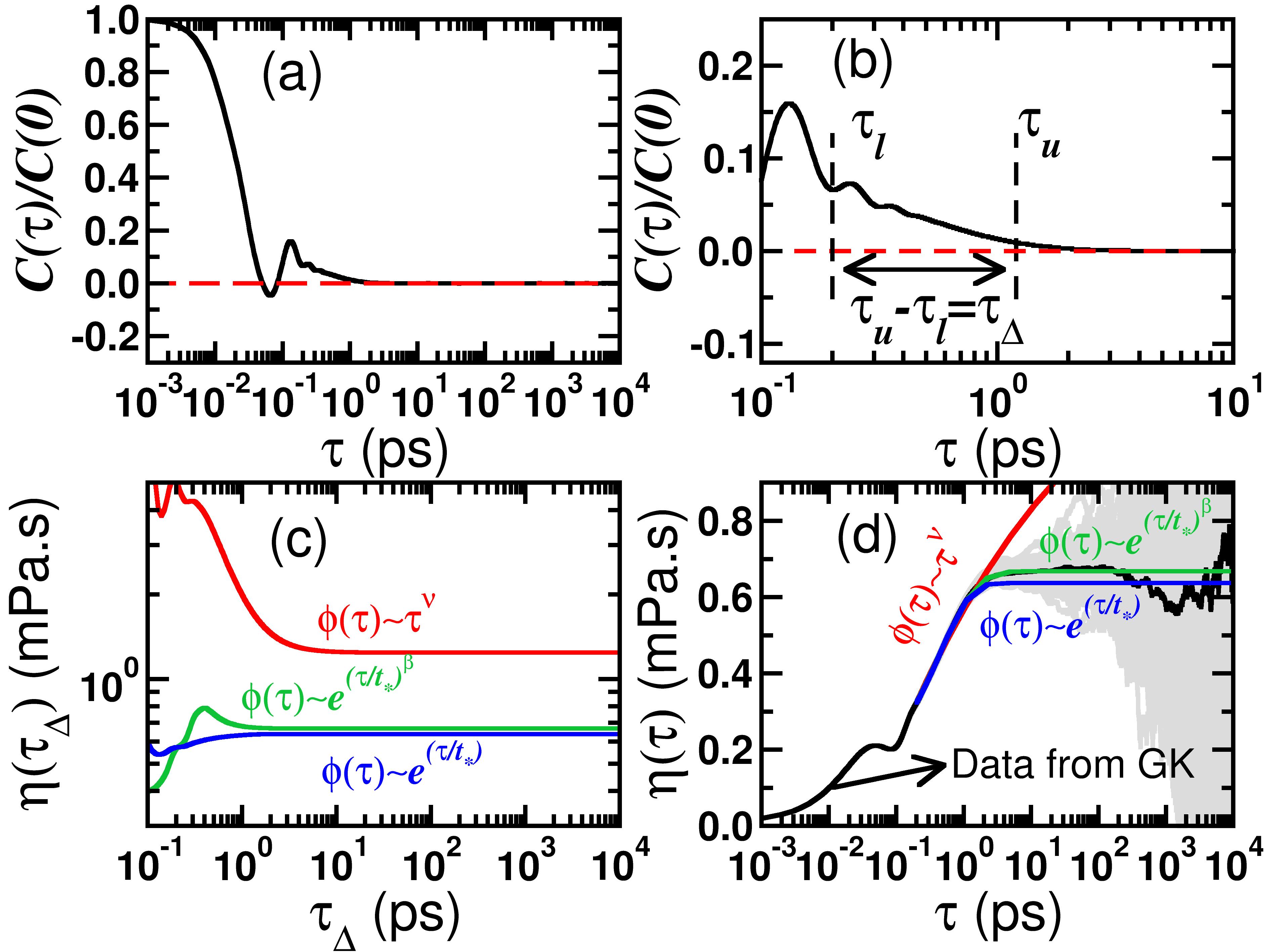}
    \caption{Benchmarking the hGK framework proposed in this Letter to SPC/E water: (a) Normalized SACF, (b) Identification of the optimal boundaries for the fitting window with width $\tau_\Delta=\tau_u-\tau_l$, (c) Predicted viscosity as a function of $\tau_\Delta$ for different analytical tail models for $\phi(\tau)$. (d) Running integral of viscosity from the GK and hGK using different tail models. Gray curves represent running integral obtained from individual trajectories, and the black curve represents their ensemble average. The hGK approach using stretched exponential yields the closest match with the GK and provides a smooth, noise free plateau.}
    \label{fig:figure2}
\end{figure}

We first evaluate the proposed hGK framework using SPC/E water, a widely investigated model liquid with well established viscosity benchmarks from both experiments and equilibrium MD simulations\cite{balasubramanian1996,gonzlez2010,fanourgakis2012}. Water is a stringent test case because its SACF shows a rapid initial decay followed by weak oscillations and a noisy long time tail, a combination that often prevents the traditional GK framework from reaching a clear plateau and thus provides an ideal benchmark for assessing the proposed hGK approach.

The SACF of SPC/E water [Figure \ref{fig:figure2}(a)] decays to near zero within a few hundred femtoseconds and then enters a weak oscillatory regime that extends to roughly 0.2 ps. Beyond this point, the SACF decreases smoothly before fluctuating around zero at longer lag times. The fluctuations at large lag times introduce noise in the running integral of viscosity using the GK framework, as seen in Figure \ref{fig:figure2}(d). Although the running integral shows a clear initial plateau, noise accumulation at large lag times obscures convergence. The hGK framework resolves this ambiguity by combining numerical integration of the short time SACF with analytical continuation of the long time tail. The lower cutoff $\tau_l$ marks the end of the fast oscillatory regime, and $\tau_u$ identifies the highest lag time used for fitting [Figure \ref{fig:figure2}(b)]. Within this interval, defined by the fitting window length $\tau_\Delta=\tau_u-\tau_l$, we fit the SACF tail to analytical models and extend the viscosity integral beyond $\tau_u$ using the fitted function $\phi(\tau)$.

To examine the sensitivity of the hGK framework to the functional form of $\phi(\tau)$, we considered three different choices for tail models: a power law decay ($\phi(\tau)\sim\tau^{-\nu}$), a simple exponential decay ($\phi(\tau)\sim\exp(-\tau/\tau_*)$), and a stretched exponential decay ($\phi(\tau)\sim\exp\left[-(\tau/\tau_*)^\beta\right]$). These forms correspond to distinct relaxation mechanisms: Power law decays appear in hydrodynamic theories of simple fluids, while exponential and stretched exponential functions describe systems with single and broad relaxation time distributions. Figure \ref{fig:figure2}(c) reports the predicted viscosity as a function of the fitting window length using different choices for tail models. The exponential and stretched exponential models yield mutually consistent viscosities, whereas the power law decay produces a significantly higher value. This discrepancy arises because the power law form misinterprets the long time contributions in SACF, which suffer from sampling in EMD.

Figure \ref{fig:figure2}(d) compares the running integral of viscosity obtained from the traditional GK framework with that of hGK for three different models of $\phi(\tau)$. While the hGK framework recovers the viscosity at low lag times for all models of $\phi(\tau)$, only the model $\phi(\tau)\sim\exp(-\tau/\tau_*)$ reproduces the results obtained from the traditional GK framework. The stretched exponential decay model of SACF at large lag times provides the most accurate prediction of viscosity, yielding a value of 0.668 [$\pm$ 0.011] mPa$\cdot$s for SPC/E water at 303 K. This value agrees closely with previous molecular dynamics studies and experiments, confirming that our hGK framework preserves the statistical mechanics foundation of the traditional GK framework while overcoming phase space sampling issues. In addition to producing smooth and stable viscosity plateaus, the hGK framework offers a computational speedup of about $10^2$ for the SPC/E system. Here, the speedup is defined relative to GK calculations performed until the running viscosity integral reaches a converged plateau, whereas the hGK framework achieves comparable estimates using substantially shorter trajectories (see Section S2 of \textbf{SI}). This gain arises from the ability to extract converged viscosities from short trajectories without requiring extensive sampling of the long time SACF tail.

After validating the hGK framework for SPC/E water, we next applied it to soft matter systems of direct relevance to rechargeable battery electrolyte technologies. Understanding viscosity in electrolytes is essential because ion transport is strongly influenced by viscous dissipation and correlated molecular motion\cite{onsager1945}. Although the classic Stokes-Einstein relation establishes an inverse relation between diffusivity and viscosity\cite{onsager1945}, the impact of viscosity extends to ion pair lifetimes, dielectric relaxation, and the extent of ion association\cite{maginn2015,mogurampelly2017,hema2025}, all of which influence device behavior. Recent studies have focused on polymer electrolytes that exhibit partial or complete decoupling of viscosity from ionic conductivity\cite{hallinan2013,ganesan2025,das2025,sharma2024}, creating promising directions for high performance solid state batteries. These emerging electrolyte chemistries introduce new challenges for theory and simulation, where slow stress relaxations render the traditional GK calculations prohibitively expensive or impractical. Efficient and reliable viscosity estimation, therefore, remains a central requirement for mechanistic understanding and materials design.

\begin{figure}[t]
    \centering
    \includegraphics[width=3.375 in, keepaspectratio]{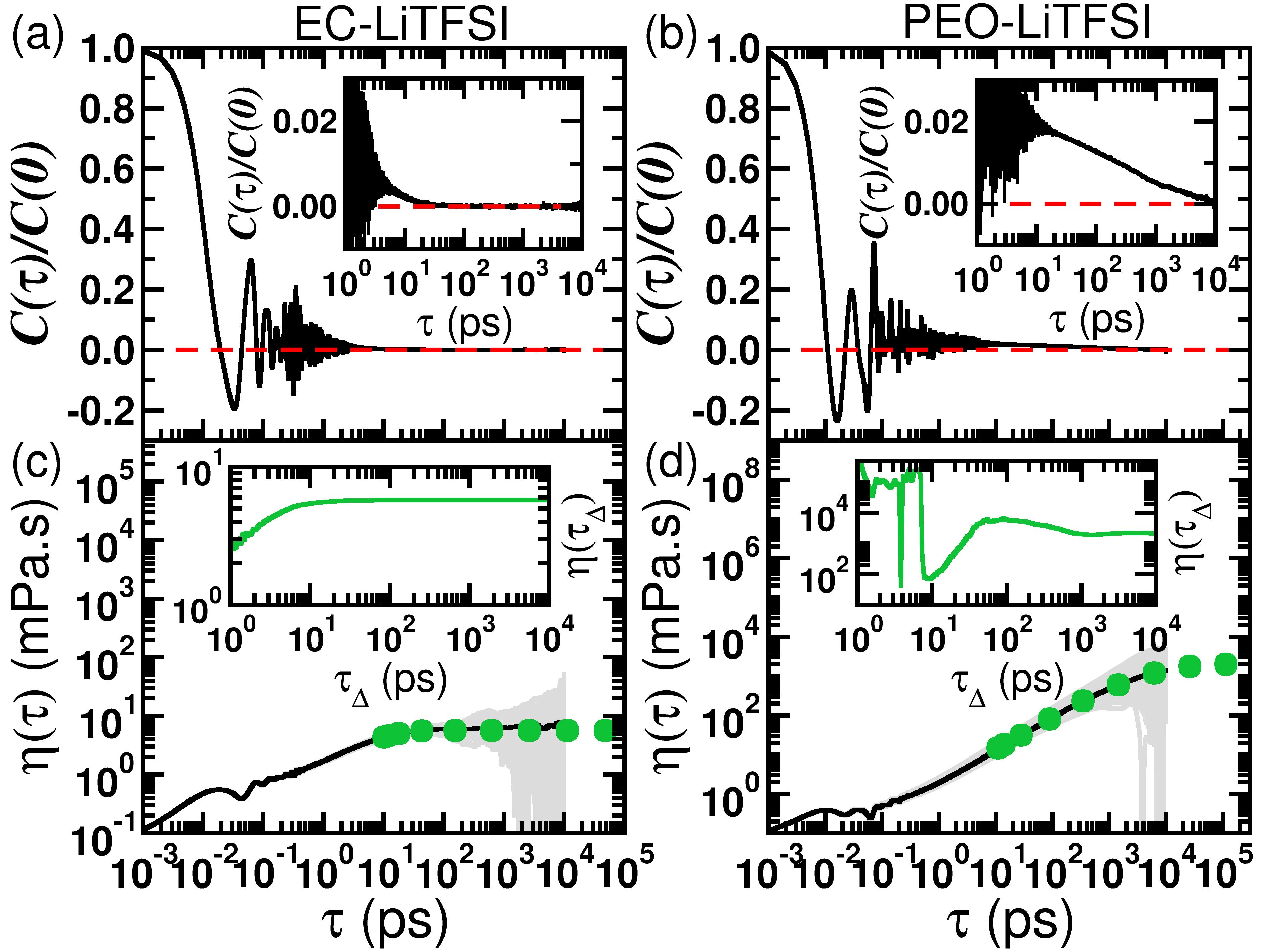}
    \caption{Demonstration of the hGK framework for liquid and polymer electrolytes: normalized SACF of (a) EC-LiTFSI and (b) PEO-LiTFSI electrolytes with slow tail highlighted in the respective insets. Running integral of viscosity for (c) EC-LiTFSI and (d) PEO-LiTFSI electrolytes. Gray curves represent running integral obtained from individual trajectories, and the black curve represents their ensemble average. The insets to the figures show the convergence behavior as a function of $\tau_\Delta$ used to fit $C(\tau)$ (see Figure \ref{fig:figure2}). The hGK framework recovers stable plateaus for both systems despite their slow relaxation and noisy long time SACF tails.}
    \label{fig:figure3}
\end{figure}

\begin{table*}[t]
\centering
\caption{System details and fitted parameters used in the hGK analysis for water, EC-LiTFSI, and PEO-LiTFSI.}
\renewcommand{\arraystretch}{1.20}
\begin{tabular}{|c|c|c|c|c|}
\hline
\multicolumn{2}{|c|}{\textbf{System}} & \textbf{SPC/E Water} & \textbf{EC-LiTFSI} & \textbf{PEO-LiTFSI} \\
\hline
\multicolumn{2}{|c|}{\#$N_\text{atoms}$} & 12165 & 24576 & 39196 \\\hline
\multicolumn{2}{|c|}{$T$ [K]} & 303 & 303 & 303 \\\hline
\multicolumn{2}{|c|}{$V$ [nm$^3$]} & 121.73 & 287.89 & 381.19 \\\hline
\multicolumn{2}{|c|}{$\tau_l$ [ps]} & 0.2 & 2 & 10 \\\hline
\multicolumn{2}{|c|}{$\tau_u$ [ps]} & 2 & 20 & 1000 \\\hline
\multicolumn{2}{|c|}{$C(0)$ [$10^{10}$ Pa$^2$]} & 33975.9 & 61935.1 & 81966.1 \\\hline
\multirow{3}{*}{%
  \parbox{3.8cm}{\centering Fitting parameters in\\[4pt]
  $\displaystyle \phi(\tau)=a_0\exp\!\Big[-\big(\tfrac{\tau}{\tau^*}\big)^{\beta}\Big]$}
}
& $a_0$            & 0.200   & 0.063   & 0.026 \\\cline{2-5}
& $\tau^{*}$ [ps]  & 0.173   & 0.319   & 243.929 \\\cline{2-5}
& $\beta$          & 0.585   & 0.350   & 0.342 \\\hline
\multicolumn{2}{|c|}{Speedup Factor, $S$ (see SI)} & $10^2$ & $10^3$ & $10^3$ \\\hline
\multicolumn{2}{|c|}{$\eta_{\mathrm{expt}}$ [mPa$\cdot$s]} & 0.79\cite{huber2009} & $\sim 5$\cite{bolloli2015} & na \\\hline
\multicolumn{2}{|c|}{$\eta_{\mathrm{GK}}$ [mPa$\cdot$s]} & $0.675 \pm 0.002$ & $5.898 \pm 0.077$ & na \\\hline
\multicolumn{2}{|c|}{$\eta_{\mathrm{hGK}}$ [mPa$\cdot$s]} & $0.668 \pm 0.011$ & $6.142 \pm 0.274$ & $1999 \pm 225$ \\\hline
\end{tabular}
\label{table1}
\end{table*}

\begin{figure*}[t]
    \centering
    \includegraphics[width=6.75 in, keepaspectratio]{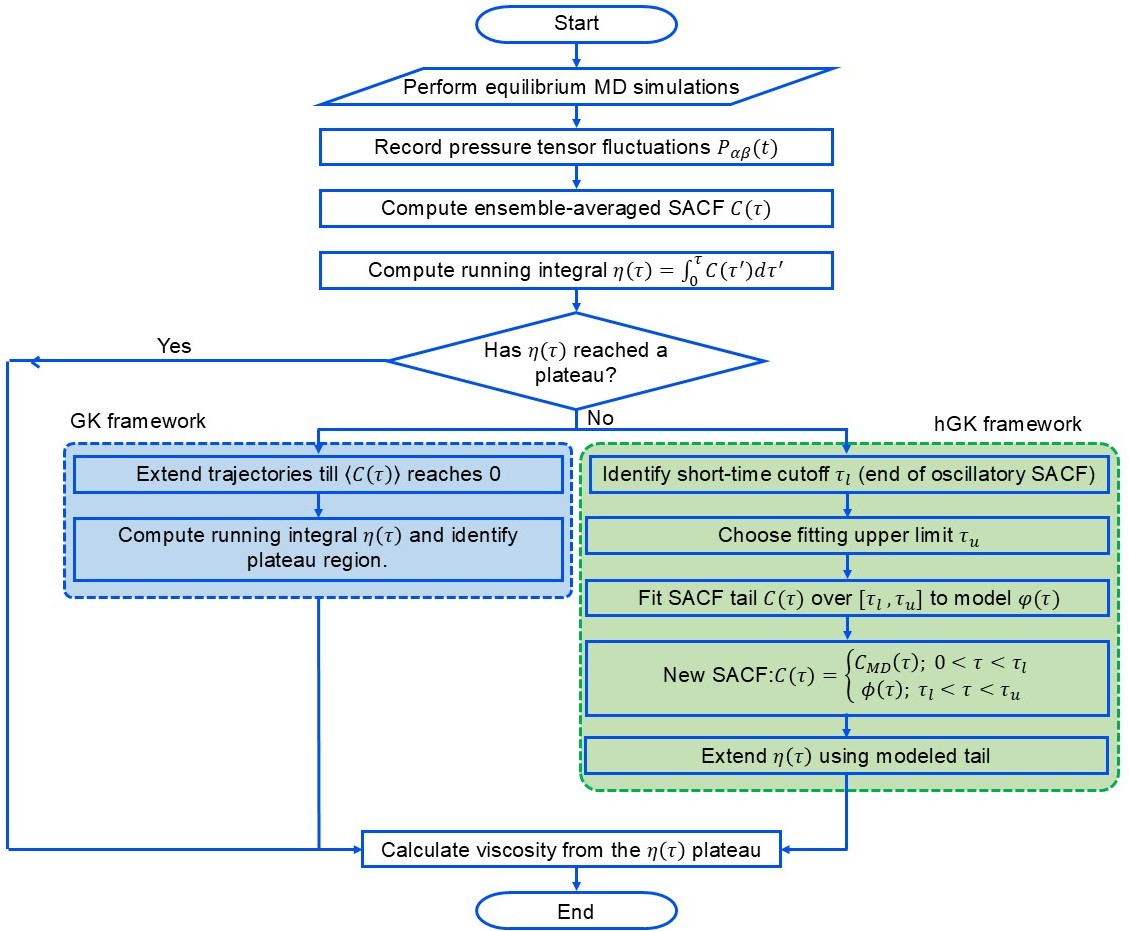}
    \caption{Workflow for viscosity calculation using the GK and hGK frameworks. The hGK protocol extracts the short time SACF from MD, models the long time tail, reconstructs the SACF, and evaluates the viscosity plateau without requiring extended trajectories.}
    \label{fig:figure4}
\end{figure*}

We examined two representative electrolytes that span distinct dynamical regimes: EC-LiTFSI, a widely used liquid electrolyte, and PEO-LiTFSI, a common polymer electrolyte. Together, these systems span a broad range of relaxation times, molecular complexity, and ion-solvent/ion-polymer coupling. The traditional GK framework requires tens of nanoseconds to microseconds of sampling for convergence in such systems, which is computationally demanding. In contrast, the hGK framework recovers converged viscosities using only $\sim$10 ns of sampled data for PEO-LiTFSI, and often much less.

Figure \ref{fig:figure3} demonstrates the performance of the hGK framework for liquid and polymer electrolytes through the analysis of their SACFs and running viscosity integrals. For EC-LiTFSI [Figure \ref{fig:figure3}(a)], the SACF decays more slowly than in water and shows a two stage relaxation that spans fast local motions and slower collective rearrangements. A rapid decay in the first few tens of femtoseconds is followed by a mild oscillatory region and a smooth tail extending beyond 50 ps (see the inset). This long time component arises from the collective structural reorganization of ion pairs and surrounding solvent molecules. In contrast, stress relaxation in PEO-LiTFSI [Figure \ref{fig:figure3}(b)] is much slower and more complex. The SACF displays persistent oscillations that extend over several picoseconds and a broad tail that decays over tens of picoseconds. Such behavior reflects the coupled segmental motion of the polymer and coordinated lithium ions, producing a wide distribution of relaxation times. When viewed on a logarithmic scale, as shown in the insets, the SACFs appear not to decay fully to zero even after 10 ns of lag time, highlighting the extreme persistence of stress correlations in this system. These SACFs highlight how increasing molecular complexity leads to extended relaxation regimes, making PEO-LiTFSI an ideal test for the robustness of the hGK framework.

We fitted the long time SACF tails using a stretched exponential form after optimizing the fitting window length $\tau_\Delta$ [see Table \ref{table1}]. This approach captures the slow decay while avoiding the noise that accumulates in the traditional GK integral, where direct numerical integration of the tail region leads to drift. The running viscosity integrals presented in Figures \ref{fig:figure3}(c-d) further illustrate these points. The running integrals from the GK framework exhibit large fluctuations at long times, leading to ambiguity in plateau identification even after ensemble averaging. For EC-LiTFSI, the viscosity reaches a well defined plateau near 6 mPa·s, in agreement with previous reports\cite{bolloli2015}. For PEO-LiTFSI, the traditional GK integral fails to converge, consistent with the presence of fast local motions and much slower collective rearrangements, whereas the hGK continuation yields a stable value near 2000 mPa$\cdot$s. The insets show that, once a minimal segment $\tau_\Delta$ has been optimized, the viscosity predicted by hGK remains essentially independent of the variations in the fitting window length (particularly $\tau_u$ for a fixed $\tau_l$), confirming the robustness of the fitted stretched exponential tail model, $\phi(\tau)$. The PEO chains in this system lie below the entanglement regime, therefore the current hGK framework is able to capture the slow relaxations of the system with just a single stretched exponential. Entangled polymer melts exhibit additional relaxation modes, resulting in a broadened relaxation spectrum\cite{mcleish2002}. Future extensions of hGK can incorporate these modes and provide a more comprehensive description of complex polymer systems. 

Together, these results demonstrate that the hGK framework captures stress relaxation across multiple timescales in both liquid and polymer electrolytes while requiring only modest simulation lengths. By combining direct integration of short time SACF data with empirically fitted continuation of the tail, the hGK framework provides a practical route for estimating viscosity in chemically and dynamically diverse electrolyte systems. This capability is essential for predictive electrolyte modeling and for understanding ion transport mechanisms in next generation battery materials.

In summary, the hGK framework provides a fast and reliable approach for predicting viscosity from equilibrium MD simulations. The method extracts the well resolved short time SACF directly from MD and replaces the poorly sampled long time tail with an analytical continuation. The hGK framework preserves the statistical mechanical foundation of the traditional GK framework while avoiding the need for prohibitively long trajectories. A schematic comparison of the traditional GK and hGK workflows is shown in Figure \ref{fig:figure4}, highlighting how the hGK framework partitions the SACF into a directly sampled short time regime and a modeled long time relaxation tail, enabling reconstruction of the viscosity plateau from short simulations. Benchmark calculations for SPC/E water confirm that hGK reproduces viscosities obtained from traditional GK and experiments while reducing the computational requirement by several orders of magnitude. Application to EC-LiTFSI and PEO-LiTFSI shows that the method captures slow ionic and polymeric relaxation processes that prevent direct GK convergence. The stretched exponential tail provides a robust and physically grounded continuation across systems that span more than four orders of magnitude in viscosity.

Several open questions remain, including the development of more principled criteria for identifying $\tau_l$ based on the molecular origins of short time stress correlations. Future studies can extend the hGK framework to incorporate the broad relaxation spectra exhibited by more complex soft matter systems. These channels include reptation, contour length changes, and constraint release. Developing automated criteria for these choices will further expand the applicability of the hGK framework. Nonetheless, the results presented here establish the hGK framework as a general, computationally efficient approach capable of treating a broad class of liquids and polymer electrolytes. Its simplicity and accuracy make it a powerful tool for predictive studies of soft materials, ionic liquids, and next generation battery electrolytes\cite{hallinan2013,marioni2025}.

\begin{center}
\noindent\textbf{Supporting Information.}
\end{center}
The data supporting this article have been included as part of the SI. The supplementary information contains: (Section S1) Applying the hGK framework to liquid Ar. (Section S2) Calculation of the computational time speedup and the speedup values for the various systems studied in this work. (Section S3) Sensitivity of the predicted viscosity on the chosen $\tau_l$. The Python scripts for the hGK framework are provided in the GitHub repository: \url{https://github.com/pemg-iitj/hGK_framework}

\begin{center}
\noindent\textbf{Acknowledgments}
\end{center}
\begin{acknowledgments}
	The authors acknowledge IIT Jodhpur for the support provided through the DGX2 and HPC. AKM acknowledges the fellowship provided by the Ministry of Education (MoE), Government of India.
\end{acknowledgments}

\bibliography{hybridGK_viscosity}
\bibliographystyle{abbrv}
\end{document}